\documentstyle[aps,epsfig,multicol]{revtex}

\newcommand{\bq}{\begin{equation}}
\newcommand{\eequ}{\end{equation}}
\newcommand{\bqa}{\begin{eqnarray}}
\newcommand{\eqa}{\end{eqnarray}}
\newcommand{\nn}{\nonumber}
\newcommand{\ms}[1]{\mbox{\scriptsize #1}}

\newcommand{\smallfrac}[2]{\mbox{$\frac{#1}{#2}$}}

\newcommand{\half}{\smallfrac{1}{2}}

\begin{document}
\draft

\title{Quantum feedback control and classical control theory}

\author{Andrew C. Doherty\footnote{Present address: 
Norman Bridge Laboratory of Physics 12-33,
California Institute of Technology, Pasadena CA 91125. Email: 
dohertya@its.caltech.edu}}
\address{Department of Physics, University of Auckland,
Auckland, Private Bag 92019, New Zealand} 
\author{Salman Habib, Kurt Jacobs}
\address{T-8, Theoretical Division, Los Alamos National
Laboratory, Los Alamos, New Mexico 87545} 
\author{Hideo Mabuchi}
\address{Norman Bridge Laboratory of Physics 12-33,
California Institute of Technology, Pasadena CA 91125}
\author{Sze M. Tan}
\address{Department of Physics, University of Auckland,
Auckland, Private Bag 92019, New Zealand} 
\maketitle

\begin{abstract}
We introduce and discuss the problem of quantum feedback control in
the context of established formulations of classical control theory,
examining conceptual analogies and essential differences. We describe
the application of state-observer based control laws, familiar in
classical control theory, to quantum systems and apply our methods to
the particular case of switching the state of a particle in a
double-well potential.
\end{abstract}


\begin{multicols}{2}

\section{Introduction}

Experimental technology, particularly in the fields of cavity
QED~\cite{CQED}, ion trapping~\cite{ion} and Bose-Einstein
condensation~\cite{andrews1996a}, has now developed to the point where
individual quantum systems can be monitored continuously with very low
noise and may be manipulated rapidly on the time-scales of the system
evolution. It is therefore natural to consider the possibility of
controlling individual quantum systems in real time using
feedback~\cite{qfb1}. In this paper we consider the problem of
feedback control at the quantum limit.  In a fully quantum mechanical
feedback control theory the quantum dynamics of the system and the
back-action of measurements must both be taken into account.

The major theoretical challenge of extending feedback control to the
quantum mechanical regime is to describe properly the back-action of
measurement on the evolution of individual quantum systems.
Fortunately, the formalism of quantum measurement, and particularly
that of the continuous observation of quantum systems, is now
sufficiently well developed to provide a general framework in which to
ask salient questions about this new subject of {\em quantum feedback
control}. In fact, the formulation that results from this theory is
sufficiently similar to that of classical control theory that the
experience gained there provides valuable insights into the
problem. However, there are also important differences which render
the quantum problem potentially more complex. In this paper we
describe a fairly general formulation of the classical feedback
control problem, and compare it with a similarly general quantum
feedback control problem. This allows us to examine ways in which the
classical problem may be mapped to the quantum problem, to provide
insight, and to show when results from the classical theory may be
applied directly to the control of quantum systems. This will also
allow us to highlight the essential features of the quantum problem
which distinguish it from classical feedback control.

The field of quantum-limited feedback was introduced 
by Wiseman and Milburn~\cite{qfb1}, who considered the instantaneous
feedback of some measured photocurrent onto the dynamics of a quantum
system. The master equation for the resulting evolution was then
Markovian. In this work we are interested in more general schemes in
which some arbitrary functional of the entire history of the
measurement results can be used to alter the system evolution. The
resulting dynamics of the system is then non-Markovian, however the
dynamics of the system and controller remain Markovian. As we shall
see this is completely analogous to the situation in classical control
theory.

The Wiseman-Milburn theory has been applied to the generation of
sub-shot noise photocurrents through feedback and the affect of the
in-loop light on the fluorescence of an
atom~\cite{taubman1995a}. Other proposed applications include the
protection and generation of non-classical states of the light
field~\cite{slosser1995a}
and the manipulation of the motional state of atoms or the mirrors of
optical cavities~\cite{dunningham1997}. In
related work Hofman {\it et al}~\cite{hofman1998a} consider the
preparation and preservation of states of a two-level atom through
homodyne detection and feedback in a slightly different
formalism. Finally the so-called `dynamical decoupling' of a quantum
system from its environment has been
discussed~\cite{viola1999a} which protects states of the
system of interest from the effects of coupling to the environment in
situations in which it is possible to manipulate the system on times
short compared to the correlation time of the environment. This is the
opposite limit to the Wiseman-Milburn theory which considers feedback
slow on the time scale of the bath correlations but fast on the
time scales of the dissipative or non-linear dynamics. This work
adopts, as we do, ideas from classical control theory, in this case
the so-called bang-bang control, to open quantum systems. There is an
extensive literature on the application of classical control
techniques such as optimal control to {\em closed} quantum systems, a
useful entry point into this literature being Ref.~\cite{warren1993a}.
  
Although the task of determining useful functionals of the measurement
current may seem daunting we argue that much progress can be made by
adopting the lessons of the classical theory of state estimation and
control. In particular it is helpful to break the feedback control
process into two steps --- the propagation of some estimate of the
state of the system given the history of the measurement results, and
the use of this state estimate at a given time in order to calculate
appropriate control inputs to affect the dynamics of the system at
that time. This approach has already yielded results for the optimal
control of observed linear quantum systems~\cite{BelavkinLQG,DJ}.

A simple example of an experiment in quantum optics in which similar
control strategies have already been employed is the work of Cohadon
{\em et al.}~\cite{CHP}. In this experiment the aim is to damp the
thermal motion of the end mirror of a high finesse optical cavity.  A
very high precision interferometric measurement is made of the
mirror's position and the resulting signal is filtered in an
appropriate way to generate an estimate of the current mirror
momentum. This momentum-estimate signal is then used to modulate the
laser power of a laser driving the back of the mirror in order to
exert a radiation pressure force in the opposite direction to the
mirror momentum, thus reducing the effective temperature of the
mirror. In fact the considerable thermal noise in the experiment means
that the back-action noise is not significant and so an essentially
classical treatment of the feedback is sufficient. In this paper we
wish to consider a relatively general description of this kind of
feedback technique in a way that explicitly takes into account the
quantum mechanical back-action noise and will thus be relevant to
experiments such as~\cite{CQED} where truly quantum control is a near
future possibility.

In the next section we describe the classical feedback control problem
well-known in classical control theory, while in Section III we
introduce a formulation of the quantum problem and examine conceptual
analogies between the two. We consider optimization of the control
strategy and discuss the quantum equivalent of the Bellman equation,
being a general statement of the quantum optimal control problem in a
dynamic programming form. In Section IV we consider the possibility of
making precise mappings between the classical and quantum problems,
and examine when the quantum problem may be addressed using the
classical theory directly. In Section V we consider the classical
concept of observability and discuss ways in which this may be defined
for quantum systems. In Section VI we consider the application of
sub-optimal control strategies developed for non-linear classical
systems to quantum systems. As an example we consider controlling the
state of a particle in a double-well potential in the presence of
noise. Section VII concludes.

\section{Classical Feedback Control}
\label{CFC}
In this section we consider the classical feedback control
problem~\cite{cfb1,Maybeck,cfb2,Ben,residual}.  It is not
our intention here to be completely general, since the control problem
is a very broad one. We will consider explicitly only continuous time
systems, and these driven by Gaussian noise. Since most of what we say
will apply also to discrete systems, and those driven by other kinds
of noise sources, little is lost by this restriction.

The problem which classical feedback control theory addresses consists
of the following: A given dynamical system, driven by noise, and
monitored imperfectly, is driven also by some input(s) with the
intention of controlling it, and these inputs are allowed to be a
function of the results of the observations performed on the
system. The dynamics of the system may be written as
\begin{equation}
d{\bf x} = {\bf F}({\bf x},{\bf u})dt + {\cal G}({\bf x},{\bf u})\cdot {\bf
dW} , 
\end{equation}
where ${\bf x}$ is the state of the system (a vector consisting of the
essential dynamical variables), ${\bf u}$ is a set of externally
controllable inputs to the system, ${\bf dW}$ is a set of Wiener
increments, and $t$ is time. Note that since ${\bf x}$ and ${\bf dW}$
are vectors, ${\cal G}$ is a matrix. In this paper we follow the terminology
of the quantum optics community and refer to the system of interest
that is to be controlled as simply the {\em system}. In the control
theory literature this is often termed the {\em process}. Hence the
noise driving the system is often referred to as the {\em process
noise}. To avoid confusion it may be useful to bear in mind that in
the control theory literature it is common to use the term system to
refer to all the parts of the control problem --- the process, the
control loop and all the noise and other inputs.  The observation
process is usually written as
\begin{equation}
  {\bf dy} = {\bf H}({\bf x},t) dt + {\cal R}(t)\cdot{\bf dV} ,
\end{equation}
where ${\bf dV}$, referred to as the {\em observation} noise, is
another set of Wiener increments which may or may not be correlated
with the noise driving the system, ${\bf dW}$.

The process of feedback control involves choosing the inputs ${\bf
u}$, at each time $t$, as some function of the entire history of the
observation process ${\bf dy}$ and of the initial conditions. To
complete the specification of a given control problem, one must define
a {\em cost function}, which specifies the desired behavior, and the
`cost' associated with deviations from this behavior. An important 
goal of control theory is then to specify ${\bf u}$ such that the cost
function is minimized. Such a result is referred to as {\em optimal
control}.

As a general principle we can say that as our knowledge regarding the
state of the system at any given time becomes better, so too does the
efficacy of the feedback algorithm, since we can better determine the
appropriate feedback. Hence the question of state-estimation (that is,
the determination of our best estimate of the state from the results
of the measurement process) arises naturally in this context. In the
fullest description, one can decide upon a probability density,
$P({\bf x})$, that describes one's complete initial state of knowledge
of the dynamical variables ${\bf x}$, and then determine how this
density evolves due to the system dynamics and the continual
observation. The equation governing this {\em a posteriori}
probability density is called the Kushner-Stratonovitch (KS) equation,
being
\bqa dP & = & -\sum_{i=1}^{n} \frac{\partial}{\partial x_i} (F_i
P) dt  \nn \\
        &   & + \frac{1}{2} \sum_{i=1}^{n} \sum_{j=1}^{n}
\frac{\partial^2}{\partial x_i \partial x_j} ([{\cal GG}^T]_{ij} P) dt \label{KSE} \\ 
& & + [{\bf H(x,t)} - \langle {\bf H}(x,t)\rangle]^T 
({\cal RR}^T)[{\bf dy} - \langle {\bf H}(x,t)\rangle dt] P. \nn
\eqa 
Here we have written the elements of ${\bf x}$ and ${\bf F}$ as
$x_i$ and $F_i$ respectively, $[{\cal GG}^T]_{ij}$ denotes the $ij^{th}$ 
element of the matrix ${\cal GG}^T$, and $\langle \ldots \rangle$ is the
expectation value with respect to $P$ at the current time.  With the
exclusion of the final term, this is merely the Fokker-Planck equation
for (unconditional) evolution of the noise-driven system. It is the
final term which takes into account the effect of the measurement on
our state of knowledge. Note that as a result of the terms involving
$\langle {\bf H}(x,t)\rangle$ this is a non-linear equation for the
probability distribution. Here we have made the usual assumption that 
the process and measurement noises are decorrelated. The stochastic 
process which drives the KS
equation is the difference between the actual measured values, ${\bf
dy}$, and the value one {\em expected} to measure, $\langle {\bf
H}(x,t)\rangle$. This is referred to as the {\em residual}, or {\em
innovation}. Since the conditioned probability distribution is the
optimal estimate of the state that may be obtained from the
measurement record, the residual has zero mean and is uncorrelated
with the conditioned probability distribution. Note that the residual
is distinct from both the process noise and the measurement noise.

It is worth mentioning that it is also possible to write a linear
equation for the conditional probability density $P$, if we relax the
requirement that $P$ be normalized. The resulting equation, which may
be found in e.g. Ref.~\cite{Ben}, is called the Zakai equation.

For linear systems driven by Gaussian noise, the KS equation becomes
particularly simple, with initially Gaussian densities remaining
Gaussian. As a result closed equations of motion for the means (being
also the `best', or maximum {\em a posteriori} estimates of the system
state) and variances can be obtained. Evolving these moment equations
is then much simpler than trying to keep track of an entire
distribution.

In addition, for linear systems the classical optimal control problem
is essentially solved. Under the assumption of a cost function
quadratic in the dynamical variables, the optimal control law involves
making ${\bf u}$ a linear function of the best estimate of the
dynamical variables, and the equation for determining this function
may be given explicitly in terms of the (in this case linear)
functions ${\bf F}$ and ${\cal G}$. Moreover, the solution of the linear
problem possesses certain important properties which make it
particularly simple: It satisfies the {\em separation theorem}, which
states that the optimal control law depends on only one estimate of
the state~\cite{cfb1,cfb2,residual} --- in this case the mean of the
{\em a posteriori} probability distribution. There is no advantage in
modifying the control law based on the uncertainty of the current
state estimate.  The linear problem also satisfies {\em certainty
equivalence}. This means that the optimal control strategy is the same
as it would be even if there was no noise driving the system and the
state of the system were known exactly; in the stochastic problem the
optimal state estimate simply takes the place of the system state in
the deterministic problem. Furthermore the linear problem is {\em
neutral}, which means that the choice of controls does not affect the
accuracy of the state estimate. If the action of the controller
affects the uncertainty about the state of the system as the well as
the evolution of the system itself this is termed {\em dual effect}.

For non-linear systems the situation is very different. Non-linear
systems may satisfy only a few of the above conditions, or none at
all. Few exact results exist for optimal control strategies. True
optimal estimation almost invariably requires the integration of the
full KS equation, something which is impractical for real-time
applications. Therefore it is generally necessary to develop good
approximate, but nevertheless sub-optimal, estimation and control
strategies, and many approaches to this problem have been
developed. In Section~\ref{subopt} we will consider similar approaches
to the quantum problem where integration of an optimal estimate of the
system state may also be impractical in real time.

Another reason for employing nominally sub-optimal feedback control is
to account for uncertainty in the model. If parameters of the model of
the system are not in fact well known then the control that is optimal
for the nominal model may in fact be a very poor control loop for
models with similar but not identical values of the parameters. This
problem can be particularly pronounced in systems with large numbers
of degrees of freedom and the solution of this problem is the domain
of {\em robust control}~\cite{RAOC}. Another control technique commonly 
used in practice is pole-placement for which quantum mechanical 
analogues could also be developed.

\section{Quantum Feedback Control}

\subsection{Continuous Quantum Measurement}
\label{sec:contqmeas}
The model of the control problem introduced above makes sense in
classical physics --- however it is implicitly assumed that it is
possible to extract information about the state of the system without
disturbing it. This is not a valid assumption in quantum mechanics,
and hence in describing any experiment on a quantum system it is
necessary to consider carefully, as well as the quantum dynamics of
the system, the coupling of the system of interest with the measuring
apparatus. To provide a similarly useful formulation of quantum
feedback control we require a model of quantum continuous measurement
with a similarly wide applicability to the classical model of the
previous section. In recent years, in the field of quantum optics,
where continuous quantum measurements are realized experimentally, a
formalism was developed to accurately describe such 
measurements~\cite{MeasP,GPZ,Carm,WMhom}, and
it was realized later that this description was identical to that
developed in the mathematical physics literature using more abstract
reasoning~\cite{MeasM,BelavkinLQG}. This formalism appears 
to fill the role for quantum systems
that the classical formulation introduced above plays for classical
systems. In order to describe noise in quantum systems we will employ
the master equation formalism and because the measurement of the
system requires some coupling to the external world the continuous
measurement of a quantum system also requires the consideration of
master equations of a particular type.

If we denote the state of the quantum system that we are concerned
with controlling as $\rho$ and the system Hamiltonian as $H$, then the
effect of measurement and environmental noise may be included by
adding two Lindblad terms to the master equation for $\rho$:
\begin{equation}
  \dot{\rho} = -i[H,\rho] + {\cal D}[Q]\rho + {\cal D}[c]\rho
\label{me1}
\end{equation}
where $D[A]\rho \equiv (2A\rho A^\dagger - A^\dagger A\rho - \rho
A^\dagger A)/2$ for an arbitrary operator $A$. When $A$ is Hermitian
this reduces to $D[A]\rho=-[A,[A,\rho]]/2$. The term $D[Q]\rho$
describes the {\em unconditional} evolution resulting from a
continuous measurement where the interaction of the measuring device
and the system is via the system operator $Q$. If $Q$ is Hermitian,
then it describes a continuous measurement of the observable
corresponding to $Q$. By unconditional evolution we mean that the
master equation describes our state of knowledge if we make the
measurement but throw away the information (the measurement record).
It is therefore the result of averaging over all the possible final
states resulting from the measurement history. Similarly, averaging
over the measurement results in the classical Kushner-Stratonovitch
equation results in a Fokker-Planck equation for the probability
distribution of the state.  The second term of the master equation,
$D[c]\rho$, describes the effect of noise due to the environment.
Since it has the same form as that of the unconditional measurement
evolution, it is always possible to view it as the result of a
measurement to which we have no access. Similarly, it is always
possible to view the measurement process as an interaction with an
environment (bath) where we are performing measurements on the bath to
obtain the information, producing a continuous measurement on the
system.

Associated with any given history of measurement results will be a
conditioned state, $\rho_{\text{c}}$, being the observer's actual
state of knowledge resulting from recording the (continuous) series of
measurement outcomes. The evolution of the conditioned state is
referred to as a quantum {\em trajectory}. If one conditions on the 
measurement of the observable $Q$, the master equation (Eq.(\ref{me1})) 
becomes~\cite{WMhom}
\begin{equation}
  \label{SME1}
  d\rho_{\text{c}}
  = dt{\cal L}_{0}\rho_{\text{c}} + dt{\cal D}[Q]\rho_{\text{c}} +
  {\cal H}[Q]\rho_{ \text{c}} dW + {\cal D}[c]\rho,
\end{equation}
which is described as a Stochastic Master Equation (SME).
Here $\cal{H}$ is defined by
\begin{equation}
  {\cal H}[\Lambda]\rho = \Lambda\rho + \rho \Lambda^\dagger - \mbox{Tr}[(\Lambda + \Lambda^\dagger)\rho]\rho.
\end{equation}
The measurement process is given in terms of the process $dW$ by
\begin{equation}
    dy = \mbox{Tr}[ \left(Q+Q^{\dagger}\right) \rho] dt + dW .
\label{dy1}
\end{equation}
Here $dW$ is a Wiener increment, and we see that there is a close
similarity between the quantum measurement process and the classical
measurement process. It should be remembered that for a fixed master
equation, it is, in fact, possible to alter ones measurements to
obtain different SME's. This is referred to as choosing a different
{\em unraveling} of the master equation. In general the SME (and
therefore the measurement process) may be driven by Poisson noise as
well as Wiener noise. We will return to this point later when we
consider feedback.

In the classical description of state estimation, it is the
conditional probability density, whose evolution is governed by the
Kushner-Stratonovitch equation, that describes the observer's complete
state of knowledge. The conditional probability density contains the
probabilities for the outcomes of all measurements which may be
performed on the system. In quantum mechanics it is the density matrix
that may be used to calculate probability distributions for arbitrary
measurements on the system. It is therefore the conditional density
matrix which replaces the conditional probability density in quantum
state estimation theory, and it is the SME which is the analogue of
the Kushner-Stratonovitch equation, being the propagator for the
optimal estimate of the quantum mechanical state given the history of
the measurement current $I_{[t_{0},t)}=\{dy(t')/dt:t_{0}\leq t'<t\}$.
Just as in the classical problem a residual process ($dW$)
uncorrelated with the state estimate arises. This zero mean noise
process is again the difference between the actual measurement result
and the result expected on the basis of previous measurements.

We also note that if one allows the conditional density matrix to be
unnormalized, it is possible to write the SME as a {\em linear}
stochastic master equation. This then, is the equivalent of the Zakai
equation of classical state-estimation, which is a linear equation
propagating an unnormalized {\em a posteriori} probability
distribution.

The SME~(\ref{SME1}), like any other master equation, may be unraveled
into trajectories of pure states obeying a stochastic evolution.  This
involves imagining that it is in fact possible to make some kind of
complete measurement on the bath and that the results of these
measurements are known to the observer.  In that case we would have
complete information about the system, so that an initial pure state
would remain pure, and we could write the stochastic master equation
instead as a Stochastic Shr\"{o}dinger Equation (SSE) for the state
vector. The result is
\bqa 
d|\psi\rangle & = & \left( -iH dt + \left[ Q-\smallfrac{1}{2}\langle
Q+Q^{\dagger}\rangle \right] dW_0 \right) |\psi\rangle \nn \\
& & + \sum_j \left(c_j-\smallfrac{1}{2}\langle
    c_j+c_{j}^{\dagger}\rangle \right) dW_j |\psi\rangle \\
& & - \smallfrac{1}{2}\left(Q^{\dagger}Q -\langle
    Q+Q^{\dagger}\rangle Q + \smallfrac{1}{4}\langle
    Q+Q^{\dagger}\rangle^2 \right)dt |\psi\rangle \nn \\ 
& & - \smallfrac{1}{2}\sum_{j} \left( c_{j}^{\dagger}c_{j} -
\langle c_j+c_{j}^{\dagger}\rangle c_j + \smallfrac{1}{4} \langle
    c_j+c_{j}^{\dagger}\rangle^2 \right) dt |\psi\rangle, \nn
\eqa
where the notation $\langle a \rangle \equiv \langle \psi | a | \psi
\rangle$ was used. Here $Q$ is once again the measured observable, and 
this time we have included an arbitrary number of noise sources, 
$c_j$, rather than merely a single noise source (determined previously 
by the operator $c$). Of the Wiener processes, $dW_0$ results from the measurement process of the real observer (measuring the observable $Q$), 
and the $dW_j$ from the fictitious measurements on the bath. Many of 
these unravelings are possible depending on what measurements are 
imagined to be performed on the bath (for example a Poisson process 
might be used, for any of the noise sources, rather than a Wiener 
process), the property that all unravelings will have in common is 
that the average of the SSE over many realizations will produce the 
correct SME. It turns out that the measurement process is now given by
\begin{equation}
    dy = \langle\psi|Q+Q^{\dagger}|\psi\rangle dt + dW_0
\label{dy2}
\end{equation}
By comparing Eqs. (\ref{dy1}) and (\ref{dy2}), we see that for a given
realization of the measurement process $dy$, since in general
$\mbox{Tr}[\left(Q+Q^{\dagger}\right)\rho] \not=
\langle\psi|Q+Q^{\dagger}|\psi\rangle$, the processes $dW$ and $dW_0$
are {\em not the same}.

Since the SSE is an equation for the state vector chosen such that the
average over all trajectories correctly reproduces the SME, the
equivalent classical object would be a stochastic equation for the
state vector ${\bf x}$ such that the average reproduced the KS
equation. Such an equation can certainly be constructed, with the
introduction of fictitious noise sources corresponding to $dW_j$ in
the SSE introduced above. The use of stochastic differential equations
to propagate Fokker-Planck equations is well known in classical
theories; the Kushner-Stratonovitch equation is simply a non-linear,
stochastic Fokker-Planck equation for the {\em a posteriori}
probability distribution. It should be noted that these fictitious
noises do not correspond to the process noise.

While we have presented quantum analogies here for many of the objects
in classical state-estimation, we have not presented analogies for the
objects that describe the underlying classical system, being the
classical state vector, process noise, and measurement noise. Such
analogies may be made at the cost of replacing the state vector,
process noise and measurement noise by operators in appropriate
Hilbert spaces. This requires the formulation of the problem in terms
of quantum stochastic differential equations (QSDE's). Space prevents
us from examining this in detail here, and the reader is referred to
the work of Gardiner {\em et al}\ for a discussion of QSDE's in the context of continuous measurement \cite{GPZ}. In Table 1 we include
the analogous quantities which result from such an analysis along with
the tentative analogies we have discussed in detail in this section.

\end{multicols}
\begin{table}
\caption{Quantum/Classical Analogies in State-Estimation}
\begin{tabular}{|l@{\hspace{0.2cm}}|l@{\hspace{0.7cm}}|}
   {\bf Classical State Estimation} & {\bf Quantum State Estimation} \\ \hline
   A Posteriori Probability Distribution & Conditioned Density Matrix \\
   Kushner-Stratonovitch Equation  & Non-linear Stochastic Master Equation \\
   Zakai Equation                  & Linear Stochastic Master Equation \\
   Innovation/Residual process     & Quantum residuals ($dy - \langle
  Q+Q^{\dagger}\rangle dt$)\\
   Fokker-Planck Equation for {\em a priori} distribution & Master Equation \\
   Fictitious noise to simulate KS Eq. using SDE & Fictitious noise to
  simulate SME using SSE \\
   State Vector & Operators for System Observables \\
   Process noise                   & Bath Noise Operators \\
   Measurement noise               & Meter Field Noise Operators\\
\end{tabular}
\label{tbl1}
\end{table}
\begin{multicols}{2}


\subsection{Controlled Quantum Systems}

The goal of feedback control of quantum systems will be to use the
continuous stream of measurement results to prepare some desired state
or enforce some desired evolution of the system. In the classical
formulation this involves effectively altering the system Hamiltonian
by adding the control inputs ${\bf u}$, which are functions of the
measurement record. Quantum mechanically the equivalent action is to
make the Hamiltonian $H$ a function of the measurement record. In an
actual experiment the variation of the Hamiltonian involves the
modulation of classical parameters such as external DC fields,
laser phases and driving strengths.

However, while feedback control of the system Hamiltonian is
sufficient to cover the full classical control problem, it is not
sufficient in the quantum case. This is because, in general, the
quantum measurement process changes the dynamics of the system.
Consequently the formulation of the full quantum feedback control
problem must also allow for the possibility that the measurement
process is also changed as a result of the observations. There are two
distinct possibilities for the modification of the measurement. The
first is to control the coupling between the system and the bath (i.e.
change the operator Q) and we might refer to this as altering the
measured observable, or altering the measurement interaction. The
second is that even for a fixed system-environment coupling one can
control the nature of the measurements made on the bath. Since in this
case the master equation describing the unconditional evolution
remains the same, but the trajectories change, we may may refer to
this as altering the measurement unraveling. Such adaptive measurements~\cite{adapt} may have distinct advantages in the 
setting of quantum control.


In a general feedback scheme, the three tools of control (the
Hamiltonian, the measured observable and the measurement unraveling)
are chosen to be some integral of the measurement record. In
particular, for state-observer based control, at each point in time
they are chosen to be a function of the best estimate of the state of
the system at that time (which is also, naturally, an integral of the
measurement record). Note that in the situation considered by Wiseman
and Milburn it is only the measurement result at the latest (most
recent) time which is used in the feedback. This
leads to various complications since the feedback must always act
after the measurement and so it is necessary to be very careful of
this ordering when deriving stochastic master equations. It is
important to note that so long as the kernel of the integral of the
measurement record is not singular and concentrated at the latest
time, these complications do not arise (for the same reason that they
do not arise in classical control theory). Certainly, the integral
required to obtain the optimal state-estimate is not singular (since
it results from integrating the SME), and this remains true in all
cases we consider here (such as the sub-optimal strategy in
Section~\ref{subopt}).

With the addition of feedback the various terms in the SME are in general
functionals of the measurement record up to the latest time $t$. 
In general, this new SME is not Markovian. However, in the special
case in which the tools of control are chosen to be a function of the
optimal state-estimate (i.e. $\rho_{\text{c}}(t)$), it follows
immediately that this SME is Markovian. Since it follows from the
Quantum Bellman equation (derived below) that the optimal control
strategy may always be achieved when using a function of the best
estimate, it follows that the optimal control strategy can always be
achieved with an SME that is Markovian. The master equation that
results from averaging over the SME trajectories however, will in
general not be Markovian. In the Wiseman-Milburn scheme even the
Markovian nature of the master equation is preserved, but that is not
the case here.

\subsection{Quantum Optimal Control: the Quantum Bellman Equation}

Classically, the optimal control problem can be written in a form
which is, at least in principle, amenable to solution via the method
of dynamic programming (to be explained below). This form is 
called the Bellman Equation, and
one can also write an equivalent quantum Bellman equation. This was
first done by Belavkin~\cite{BelBell,BelavkinLQG,Belavkin99}, but since the
treatment in~\cite{BelBell} is very abstract, and since neither
optimization over unravelings, nor the possibility of ensemble
dependent cost functions were mentioned there, we feel it worthwhile
deriving this equation here using a simpler, although less rigorous
method.

To define an optimal control problem we must specify a cost function
$f(\rho (t),u(t),t)$, which defines how far the system is from the
desired state, how much this `costs', and how much a given control
`costs' to implement. The problem then involves finding the control
which minimizes the value of the cost function integrated over the
time during which the control is acting. The important point to note
is that the cost function can almost always be written as a function
of the conditional density matrix followed by an average over
trajectories. This is because the density matrix determines completely
the probabilities of all future measurements that can be made on the
system, and consequently captures completely the future behavior of
the system as far as future observers are concerned (given that the
dynamics are known, of course), which is what one almost always wants
to control.

The possible exceptions to this rule come when one is interested in
 preserving or manipulating unknown information which has been encoded 
in the system by a previous observer who prepared it in one of a known 
ensemble of states. Thus as far as the second observer is concerned the 
state of the system is found by averaging over these states with the 
weighting appropriate to the ensemble. However, in this case it may 
well be sensible to use a cost function that depends on the ensemble 
as well as this density matrix~\cite{fuchs}. It remains a topic for 
future work to determine whether problems such as this will constitute 
an important application of quantum feedback control. We will restrict 
ourselves here to what might be referred to as `orthodox' control 
objectives in which it is only the future behavior of the system which 
is important, and this is captured by cost functions which depend only 
on the density matrix (ensemble independent cost functions).

The general statement of our optimal control problem may therefore be
written as
\begin{eqnarray}
     {\cal C} = \left< \int_0^T f(\rho_{\text{c}}(t),u(t),t) dt +
     f_{\text{f}}(\rho_{\text{c}}(T),T) \right> . 
\end{eqnarray}
Here ${\cal C}$ denotes the total average cost for a given control
strategy $u(t)$, $f$ is the cost function up until the final time $T$, $f_{\text{f}}$ is the cost function associated with the final state, 
and $\langle\ldots\rangle$
denotes the average over all trajectories. The solution is given by
minimizing ${\cal C}$ over $u(t)$, to obtain the minimal cost ${\cal
C}^*$, and resulting optimal strategy $u^*(t)$. Note that the values
of $u$ will be different for different trajectories. In this
formulation a cost is specified at each point in time, with the total
cost merely the integral over time, and an allowance is explicitly
made for extra weighting to be given to the cost of the state at the
final time. It is crucial that the cost function takes this `local in
time' form in order that it be rewritten as a Bellman equation.

To derive the quantum Bellman equation we will consider the problem to
be discrete in time, since this provides the clearest treatment.  In
any case the continuous limit may be taken at the end of the
derivation, if the result is desired. In this case, dividing the
interval $[0,T]$ into N steps, the cost function consists of a sum of
the costs at times $t_i=t_1,\ldots,t_{N+1}$, with $t_{N+1}=T$ denoting
the final time. The idea of dynamic programming (which results from 
the Bellman equation) is that if the period
of control is broken into two steps, then the optimal control during
the second step must be the control that would be chosen by optimizing
over the later time period alone given the initial state reached after
the first step. This allows the optimal control to be calculated from
a recursion relation that runs backwards from the final time, or in the
continuous-time case from a backwards time differential equation. To 
derive the Bellman equation one proceeds as follows.

Trivially, at the final time, given the state $\rho(T)$, the minimal
cost is merely the final cost, so ${\cal C}^*(t_{N+1}) =
f_{\text{f}}(\rho(T),T)$. Next, stepping back to the time $t_{N}$, 
the total cost-to-go, given the state $\rho(t_N)$ is
\begin{eqnarray}
     {\cal C}(t_N) & = & f(\rho_{\text{c}}(t_N),u(t_N),t_N) \Delta t \\
& + & \int f_{\text{f}}(\rho(T),T)
P_{\ms{c}}(\rho(T)|\rho_{\text{c}}(t_N),u(t_N)) d\rho(T) \nn 
\end{eqnarray}
where $P_{\ms{c}}$ is the conditional probability density for the
state at time $T$ given the state $\rho_{\text{c}}(t_N)$, which is
conditioned on any earlier measurement results and controls, and the
control $u(t_N)$ at time $t_N$, so that the integral is simply the
conditional expectation value of the cost at the final time. Note that
the choice of the control $u(t_N)$ may depend on the measurement
result at $t_N$ and that the conditional probability density is
conditioned not only on the chosen value of $u(t_N)$ but also on the
measurement result at $t_N$. Since, $f_{\text{f}}(\rho(T),T)$ is ${\cal
C}(t_{N+1})$, we have
\begin{eqnarray}
     {\cal C}(t_N) & = & \min_{u(t_N)} \biggl[
                      f(\rho(t_N)_{\text{c}},u(t_N),t_N) \Delta t
                      \biggr.\label{bell1}  \\
                      & + & \biggl. \int {\cal C}(t_{N+1}) P_{\ms{c}} (\rho(t_{N+1})|\rho(t_N)_{\text{c}},u(t_N)) d\rho(t_{N+1}) \biggr] \nn
\end{eqnarray}
The important step comes when we consider the total cost-to-go at the
third-to-last time $t_{N-1}$. This time there are three terms in the
sum. Nevertheless, using the Chapman-Kolmogorov equation for the
conditional probability densities, it is straightforward to write the
equation for ${\cal C}(t_{N-1})$ in {\em precisely} the same form as
that for ${\cal C}(t_N)$: it is simply Eq.~(\ref{bell1}) with $N$
replaced with $N-1$. In fact, this equation holds for {\em every}
${\cal C}(t_i), i=1,\ldots,N$.

From this point, the crucial fact that results in the Bellman equation
is this: since the conditional probability densities are positive
definite, it follows that the minimum of ${\cal C}(t_i)$ is only
obtained by choosing ${\cal C}(t_{i+1})$ to be minimum. We can
therefore write a backwards-in-time recursion relation for the minimum
cost, being 
\begin{eqnarray} {\cal C}^*(t_i) & = & \min_{u(t_i)}
\biggl[ f(\rho_{\text{c}}(t_i),u(t_i),t_i) \Delta t
\biggr. \label{bellq} \\ & + & \biggl. \int {\cal C}^*(t_{i+1})
P_{\ms{c}} (\rho(t_{i+1})|\rho_{\text{c}}(t_i),u(t_i)) d\rho(t_{i+1})
\biggr] \nn ,
\end{eqnarray}
which is the discrete time version of the Bellman equation. In words,
this states that an optimal strategy has the property that, whatever
any initial states and decisions, all remaining decisions must
constitute an optimal strategy with regard to the state that results
from the first decision, which is referred to as the `optimality
principle'. 

The quantum Bellman equation confirms the intuitive result that any
optimal quantum control strategy concerned only with the future
behavior of the system is a function only of the conditional density
matrix, and further, that the strategy at time $t$ is only a function
of the conditioned density matrix at that time.

The procedure of stepping back through successive time steps from the
final time to obtain the optimal strategy is referred to as dynamic
programming. This could be used, at least in principle, to solve the
problem numerically. In practice it will be useful to employ some
approximate strategy. Much progress in this direction has been made
for closed quantum systems, see for example Ref.~\cite{botina1997a}.



\section{Classical Analogies for the Quantum Control Problem}

In the preceding sections we have examined the conceptual mappings
between the elements of the classical and quantum control problems. In
this section we want to examine the possibility of making such a
mapping precise. That is, to address the question of if and when it is
possible to model a given quantum control problem exactly as a
classical control problem. When this is possible it allows the quantum
problem in question to be solved using the relevant classical methods.

One can always formulate a given quantum control problem using the
quantum Bellman equation, but the different cost functions will be
motivated by different control objectives, and to formulate an
equivalent classical control problem we should examine these objects
of control. For example, as the object of control one might focus on
the expectation values of a set of observables, the state-vector of
the quantum system, or the entire set of density matrix elements
describing one's state of knowledge. Once we have a vector of
quantities to control, we can ask whether, if we identify this set of
quantities with the classical object of control (being the system
state vector ${\bf x}$), there exists an identical classical control
problem. In what follows we examine when this can be achieved for the
three objects of control we have mentioned.

\subsection{Correspondence using physical observables}

In this case we wish to control a vector consisting of the expectation
values of a set of observables (or, more precisely, the {\em
conditional} expectation values of a set of observables). To formulate
an equivalent classical problem we identify these with the conditional
expectation values of the classical vector ${\bf x}$, and ask whether
there exists a classical problem corresponding to a given quantum
problem. It is immediately clear that in general there will not be,
because the conditional joint probability density (e.g.\ the Wigner
function) for the quantum observables will in general not be positive
definite, while the classical equivalent is forced to be.  However, it
turns out that whenever both the quantum dynamics and the measurement
is linear in the observables, and the measurement process (unraveling)
is Gaussian, there exists an identical linear classical problem driven
by Gaussian noise, and therefore the quantum problem reduces to a
classical one. This is possible because in this case the quantum
dynamics preserves the positivity of the joint conditional probability
density.

The simplest example of this is the quantum single particle in a
quadratic potential. The equivalent classical control problem is that
for a single classical particle subject to the same potential, driven
by Gaussian noise, and with an imperfect measurement on whatever
observable is being measured in the quantum problem. Because it is the
expectation values of quantum observables which correspond physically
with the classical dynamical variables ${\bf x}$, we can denote this
formulation as using a physical correspondence between the quantum and
classical systems. Because the equivalent classical problem is linear,
it provides immediately an analytic solution to the optimal quantum
control problem for those cost functions for which solutions have been
found for the classical problem. Solutions exist for cost functions
that are quadratic in the classical variables (the so-called
Linear-Quadratic-Gaussian (LQG) theory) and also those exponential in
the variables (Linear-Exponential-Gaussian (LEG) theory). A detailed
treatment of this analogy, and the resulting quantum LQG theory is
given in Ref.~\cite{DJ}, and a rigorous mathematical treatment using a
different approach may be found in Ref.~\cite{BelavkinLQG}.

An interesting feature of this quantum-classical control analogy is
that for {\em non-linear} quantum systems it transforms smoothly from
a quantum control problem (not amenable to a classical formulation) to
a classical control problem across the quantum-to-classical
transition: from a number of numerical studies, it is now clear that
continuously observed quantum systems behave as classical systems in
the classical regime (even in the absence of any source of decoherence
other than the measurement process)~\cite{Percival}. By the
classical regime we mean the regime in which macroscopic objects
exist, with $\hbar$ small compared to the classical action, and this
therefore provides an explanation for the emergence of classical
mechanics from quantum mechanics. This has an immediate connection to
the problem of feedback control in quantum systems since feedback
controlled systems are observed systems (and the ones we are
interested in here are continuously observed). Since it is the
expectation values of the physical observables which behave as the
classical observables in the classical regime, in this regime the
above procedure will provide an effective equivalent classical control
problem. Effective non-linear classical control strategies will
therefore work in the classical regime, and a natural question to ask
is then how they perform as the system makes the quantum-to-classical
transition, and especially, whether such classical control strategies
will still work deep in the quantum regime. We explore this question
in Section~\ref{subopt}.

\subsection{Correspondence using the quantum state vector}

In this case it is the quantum state-vector $|\psi\rangle$ which is
the object of control, and so we wish to see whether we can form an
equivalent classical problem with ${\bf x}$ identified as the state
$|\psi\rangle$. In the classical case our state of knowledge is
described by the probability density, $P({\bf x})$, so that in order
to pursue a classical formulation we must consider a probability
density over the states, $P_{\ms{q}}(|\psi\rangle)$. However, there
are important differences between the roles of $P$ and $P_q$. While in
the classical case a complete knowledge of $P$ is required to predict
the results of measurements performed on the system, in the quantum
case it is only the density matrix which is required, being the set of
second moments of $P_q$:
\begin{equation}
  \rho = \int d|\psi\rangle P_{\ms{q}}(|\psi\rangle)
   |\psi\rangle\langle\psi| ,
\end{equation}
Two important consequences of this are the following. First, that
because it is only the set of second moments that characterize our
state of knowledge, many different densities $P_q$ may be chosen to
correspond to this state of knowledge, and in particular, these can
have different modes or means. Since the classical best estimate is
usually defined as a mode (maximum a posteriori estimator) or a mean,
we must immediately conclude that there is no quantum `best estimate'
for the state vector in the classical sense. Referring back to Section
\ref{CFC} then, it follows that there are no separable quantum control
problems when it is the state-vector that is the object of control.
Nevertheless, this does not rule out the possibility that it might be
useful to construct definitions of quantum `best estimates' for the
state vector in the development of sub-optimal control laws.

Second, because the equation which propagates our state of knowledge
is an equation for the density matrix, the quantum problem
automatically has moment closure. In general, the term moment closure
means that the equation for the evolution of some finite set of
moments of the conditional probability density can be written only in
terms of themselves, without coupling to the infinite set of higher
moments. In a sense, this fact introduces a simplification into the
quantum problem.

To obtain a classical model one requires that there exists a noise
driven classical system, with state vector ${\bf x}$, such that the
equation of motion for ${\bf x}$, along with the continuous
observation, whatever it may be, gives a conditional probability
density, the second moments of which obey the quantum SME. We now
present strong evidence to suggest that this is, in fact, not
possible. That is, there exists {\em no} observed classical system
that reproduces the SME, and consequently it is not possible to think
of the quantum measurement process as a classical estimation process
on the state vector. Note that this is not directly connected to the
Heisenberg uncertainty principle: the quantum state vector can be
determined completely during the observation process, just as can the
classical state. Nevertheless, the processes are fundamentally
different.

To see this first consider the equation for the second moments that
results from the the KS equation (Eq.~(\ref{KSE})), for time 
invariant linear observations on a time invariant linear system. 
In this case ${\bf F} = {\cal F}{\bf x}$, 
${\bf H} = {\cal H}{\bf x}$  and ${\cal F}$, 
${\cal H}$, ${\cal G}$ and ${\cal R}$ are constant matrices. 
The equation for the second moments may be written
\bqa
dC & = & [C {\cal F}^\dagger + {\cal F} C] - {\cal GG}^\dagger dt + \\
& + & \langle {\bf x} \rangle {\bf dW}^\dagger \sqrt{{\cal RR}^T}{\cal H}
(C - \langle {\bf x}\rangle \langle {\bf x}^\dagger\rangle) \nn \\
& + & (C - \langle {\bf x}\rangle \langle {\bf x}^\dagger\rangle)
{\cal H}^\dagger \sqrt{{\cal RR}^T} {\bf dW} \langle {\bf x}^\dagger\rangle \nn 
\eqa
where $C = \langle {\bf x} {\bf x}^\dagger\rangle$ is the matrix of
second moments. While the terms involving $F$ reproduce the commutator
for the Hamiltonian evolution of the density matrix (with the choice
$F = -iH$), as expected, the deterministic and stochastic terms
resulting from the observation are quite different. In particular, the
deterministic part is constant (i.e. not a function of $C$), and the
stochastic part depends upon the first moments. The first moments
themselves obey a stochastic equation, where the deterministic part is
given by $F$. We therefore cannot choose a linear classical estimation
problem directly equivalent to the quantum problem. If we consider
classical systems with non-linear deterministic dynamics, then the
deterministic motion fails to match the quantum evolution, which is
strictly linear. If one chooses the noise or the measurement process
to be non-linear, then, in general, the moment closure is lost.

We can gain some insight into the difference between quantum and
classical estimation by considering the change in the quantum
probability density, $P(|\psi\rangle)$, upon the result of a
measurement. Given a measurement described by the POVM $\sum\Omega_y
\Omega_y^\dagger$, and an initial density matrix $\rho$, the
post-measurement density matrix is given by $\rho' = \Omega_y \rho
\Omega_y^\dagger/{\rm Tr}(\rho \Sigma_y^{\dagger}\Sigma_y)$. Writing
this in terms of $P(|\psi\rangle)$, we have the post measurement
density for result $y$ as
\begin{equation}
  P'(|\psi_y\rangle)
      = \frac{1}{N}P(y||\psi\rangle)P(|\psi\rangle) \left|
        \frac{d|\psi_y\rangle}{d|\psi\rangle}\right| \end{equation} where \begin{equation}
        |\psi_y\rangle = \frac{\Omega_y|\psi\rangle}{\sqrt{\langle \psi
        |\Omega_y^\dagger\Omega_y|\psi \rangle}}
\end{equation}
and $P(y||\psi\rangle)$ is the conditional probability for the result
$y$ given the state $|\psi\rangle$, with $N$ a normalization. In
contrast to this, the classical result is simply Bayes' rule, being
\begin{equation}
  P'({\bf x}) = \frac{1}{N} P({\bf x})P(y|{\bf x})
\end{equation}

We see that the quantum result is Bayes rule, with the addition of a
non-linear transformation of the states, since if we set
$|\psi_y\rangle = |\psi\rangle$ for all $|\psi\rangle$ in the quantum
rule, we recover the classical Bayes rule. This is the sense in which
we can view the quantum measurement process as an active process,
since it is equivalent to a classical (passive) measurement process,
with the addition of an (active) transformation of the states.

\subsection{Correspondence using the density matrix}

In this case one considers the elements of the (conditional) density
matrix as the vector to control. Since the density matrix
characterizes our state of knowledge, by definition we always know
what it is. Consequently the SME becomes the fundamental dynamical
equation, and there is no longer any estimation in the control
problem. This is exactly analogous to considering the conditional
probability density of the classical control problem as the object of
control. Since there is no estimation the control problem is
automatically a classical one, and all the techniques of classical
control theory can be applied. However, the problem is necessarily
non-linear since the SME is non-linear.

\section{Observability and controllability}

Observability and controllability are two key concepts in classical
control theory, and here we want to examine ways in which they may be
extended to the quantum domain. They are useful because they indicate
the existence of absolute limits to observation and control in some
systems. If it is not possible to completely determine the state of a
system given a chosen measurement or to prepare an arbitrary state of
the system given the chosen control Hamiltonian then this will place
severe limitations on the feedback control of that system. It is
important to note that the definitions of observability and
controllability apply classically to noiseless systems (that is,
systems with neither process nor measurement noise), although they are
relevant for stochastic systems, and it is these systems in which we
are naturally interested here.

Consider the concept of observability. A system is defined to be
observable if the initial state of the system can be determined from
the time history of the output (i.e. the measurements made on the
system from the initial time onwards)~\cite{RAOC}. It follows that in
an observable system, {\em every} element in the (classical)
state-vector affects at least one element in the output vector, so
that the relation can be inverted to obtain the initial state from the
outputs. If one considers adding process and measurement noise, then
observability is still a useful concept, because it tells us that the
outputs, while corrupted by noise, nevertheless provide information
about {\em every} element in the state-vector. Consequently, given
imprecise initial knowledge of the state, we can expect our knowledge
of all the elements to improve with time. For an unobservable system,
there will be at least one state element about which the measurement
provides no information. The simplest example of this is a free
particle in which the momentum is observed. Since the position never
affects the momentum, any initial uncertainty in the position will not
be reduced by the measurement. Note that observability is a joint
property of a system and the kind of measurement that is being made
upon it.

It is interesting that there are at least two inequivalent ways in
which this concept of observability may be applied to a measured
quantum system, and these result from the choice of making an analogy
either in terms of the quantum state-vector, or a set of quantum
observables.  First consider observability defined in terms of a set
of observables.  The concept of observability applies in this case to
whether or not the output contains information about all the physical
observables in question. A simple example once again consists of the
single particle, in which we can use the position and momentum as the
relevant set of observables. If we consider the observation of the
position, then the system is observable: the output contains
information about both the position and momentum since the momentum
continually affects the position. As a result a large initial
uncertainty in both variables is reduced during the
observation. Naturally this is eventually limited by the uncertainty
principle. The conditioned state may eventually become pure but there
will be a finite limiting variance in the measured quantity since this
state must obey the uncertainty relations. In linear systems the
measurement back action noise has a role rather similar to process
noise in a classical system since process noise also leads to non-zero
limiting variances of the measured property of the state. This kind of
behavior is discussed in Ref.~\cite{purity}.

If we consider alternatively the measurement of momentum on a quantum
free particle, the system is unobservable, in exactly the same fashion
as the classical system is unobservable, since the momentum provides
no information about the position. It is not entirely coincidental
that in quantum mechanics momentum is a Quantum Non-Demolition (QND)
observable of the free particle while classically momentum measurement
of a free particle does not constitute an observable system. This is
clearly a general result: when it is a QND observable that is
observed, the system is always unobservable. This follows from the
fact that a QND observable is defined as one that commutes with the
Hamiltonian. Since it commutes with the Hamiltonian, no other system
observable can appear in its equation of motion, with the result that
its observation can provide no information about any other
observable. There will however be measurements on systems which while they are not
classically observable are also not QND measurements.

An alternative way to define quantum observability is in terms of the
state-vector. In this case the question of observability concerns
whether or not the output contains information about all the elements
of the quantum state vector. Consider a quantum system in which the
observation is the only source of noise. Then, if the system is
observable with respect to a particular measurement, as time proceeds
one obtains increasingly more information about all the elements of
the state vector, and the conditioned state tends to a pure state as
$t\rightarrow\infty$. For an unobservable system, any initial
uncertainty in at least one state vector element remains, even in the
long time limit. A simple example of a system that is observable in
this sense is the measurement of momentum on a free particle (recall
that this is {\em unobservable} in the previous sense). In this case
it is a simple matter to calculate the time evolution of the purity of the conditioned state
(using, for example, the method in Ref.~\cite{EvOp}), to verify that
the system is observable. An example of an unobservable system is a
set of two non-interacting spins, in which it is an observable of only
one of the spins that is measured. In this case, while the state of
the measured spin may become pure, clearly the state of the joint
system can remain mixed for a suitable choice of initial state.

A key factor which differs between these examples is that in the
observable case the measured quantity (being the momentum) has a
non-degenerate eigenspectrum, whereas in the unobservable case the
measured quantity (being any observable of the first spin) has
degenerate eigenvalues when written as an operator on the full (two
spin) system. It is clear that in the case that the measured
observable commutes with the system Hamiltonian the non-degeneracy of
the eigenvalues of the observable is a necessary and sufficient
condition for observability in this sense. Writing the evolution of
the system as multiplication by a series of measurement operators
alternating with unitary operators (due to the Hamiltonian evolution),
the measurement operators may be combined together since they commute
with the unitary operators, and it is readily shown that as
$t\rightarrow\infty$, one is left with a projection onto the basis of
the measured observable. If the eigenvalues of the observable are all
different, then the measurement results distinguish the resulting
eigenvector, and the result is a pure state. However, if any two of
the eigenvectors are degenerate, the measurement results will not
distinguish those two states.  Consequently, if the system exists
initially in a mixture of these two states it will remain so for all
time. Whether this continues to be true in the general case remains an
open question.


We need not consider controllability in any detail here, since this
has been considered elsewhere. The controllability of quantum
mechanical systems --- that is, whether the interaction Hamiltonians
available are able to prepare an arbitrary state of a quantum system
--- has been considered by applying directly the ideas of classical
control theory \cite{huang1983a}.  Interestingly, this has a new
interpretation in quantum computation.  The gates of the computer must
be able to perform an arbitrary unitary operation on the register of
qubits; a set of gates with this property is termed universal. Since
it may perform arbitrary unitary operations a universal quantum
computer may prepare any desired state of the system from any given
initial state. The conditions for controllability of a quantum system
were therefore rediscovered as the conditions for universality of a
quantum computer \cite {lloyd1995a}.


\section{Sub-optimal estimation and control for a non-linear quantum
system} 
\label{subopt}

Here we examine the application of sub-optimal estimation and control
laws, developed for non-linear classical systems, to the corresponding
quantum systems, where the objects of control are the expectation
values of physical observables. This gives a simple initial example of
the use of state observer based control systems outside of the regime
of linear systems considered in Ref.~\cite{DJ}. Since, for this
particular control objective, it is possible to completely solve the
problem of the feedback control of linear quantum systems using
classical methods for linear systems, and since continuously observed
non-linear quantum systems in the classical regime are clearly
amenable to classical control strategies, it remains to examine the
effectiveness of classical non-linear control strategies for quantum
systems deep in the quantum regime.  For non-linear systems, optimal
estimation involves integration of the KS equation for classical
systems, and the SME for quantum systems.  For real time control this
is almost always computationally impractical, so that it is important
to develop simpler (sub-optimal) algorithms which are sufficiently
accurate.

It is important to note that the use of a sub-optimal estimation
algorithm also makes the task of simulating the controlled quantum
system computationally less expensive. This is because it allows the
system, including control, to be simulated using an SSE rather than
the full SME. The reason for this is that regardless of whether the
observer is dynamically changing the inputs to the system the SSE
correctly simulates the SME --- the full SME need only be integrated
if the actual conditioned state is required to calculate the sequence
of controls. As a result, to simulate a controlled quantum system, one
need only integrate the sub-optimal estimator, if one is available,
and the SSE for the system.

Here we use as an example system a particle in a double well potential
with the control objective of keeping the particle in a given well,
and switching it from one well to the other when desired, in the
presence of a coupling to an (infinitely) high temperature bath. As
discussed in previous sections, the first important choice in such a
problem is that of the measurement, as this should be chosen so as not
to cause any unwanted dynamics (i.e. it should not force the particle
away from the desired states) and since it is the position of the
particle that is to be controlled, a position measurement is a
sensible choice.

Various approximate estimators have been developed for classical
systems, and these usually involve a moment truncation of the KS
equation. For example, one can assume that the conditional probability
density will remain Gaussian, and truncate the moments accordingly.
More generally, for a given control problem certain characteristics of
the conditional probability density might be known, and motivate
another approximation. In both the classical and the quantum
mechanical systems it is a reasonable expectation that the conditioned
states will remain Gaussian for sufficiently strong position
measurement which is the regime we will investigate here.

For the purposes of feedback control we will assume that the observer
has the ability to apply a linear force to the double well, so the
feedback Hamiltonian is proportional to $x$. When the quantum state is
close to Gaussian, quantum dynamics follows closely the equivalent
classical dynamics, and we can expect non-linear classical control
strategies to work. The strategy we will apply is that of linearized
LQG optimal control. In this method, for each time-step, the system
dynamics are linearized about the current state-estimate, and the
corresponding optimal LQG strategy is chosen for the next
time-step. In this way the control is always `locally optimal'.
Clearly the key requirement for the strategy we have outlined is that
the conditioned state remains closely Gaussian during the
evolution. The control will fail if the measurement fails to maintain
the Gaussian distribution, or if the measurement only maintains a
Gaussian at the expense of introducing an intolerable amount of noise.

The Hamiltonian for the system is
\begin{equation}
  H = \half p^2 - A x^2 + B x^4,
\end{equation}
where we have set the particle mass to unity. We will also use
$\hbar=1$. The resulting SME is
\begin{eqnarray}
  d\rho_{\text{c}} & = & -i[H + H_{\ms{fb}},\rho_{\text{c}}]dt
   + 2\beta {\cal D}[x]\rho_{\text{c}}dt \nonumber \\
  &  & + 2k{\cal D}[x]\rho_{\text{c}}dt +
   \sqrt{2k}{\cal H}[x]\rho_{ \text{c}}dW. \label{twellsme}
\end{eqnarray}
where $k$ gives the strength of the position measurement, and $\beta$
the strength of the thermal noise. On any given trajectory the
corresponding measured current is $I(t)=dQ(t)/dt$ where $dQ(t) =
\text{Tr}(x\rho_{\text{c}}(t))+dW(t)$. The feedback Hamiltonian is
$H_{\ms{fb}} = -ux$ where $u$ is a function of the history of the
photocurrent described below.

\begin{figure}
\centerline{\psfig{file=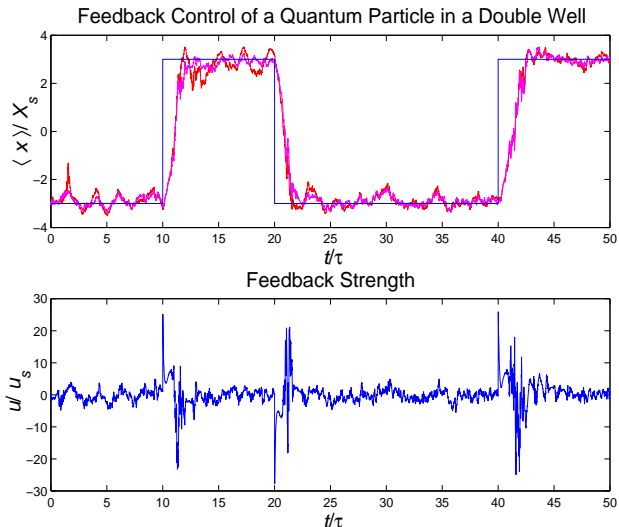,width=3.25in,height=2.8in}}
\caption{\narrowtext Behavior of a particle under the
estimation/feedback control scheme outlined in the text. (a) The
target position (blue line), the `true' mean position obtained from
the SSE simulation (red line), and the estimated position (magenta
line). (b) The control strength (size of applied force) as a function
of time. The various units are $X_{\mbox{\scriptsize s}} = 
\sqrt{\hbar/(m\nu)}$, $u_{\mbox{\scriptsize s}} = \nu\sqrt{\hbar m\nu}$ 
and $\tau = 1/\nu$, where $m$ is the mass of the 
particle and $\nu$ is an arbitrary frequency. In the text we have set 
$\hbar = m = \nu = 1$, so that all quantities are dimensionless.}
\label{fig1}
\end{figure}

The estimator chosen is a variational solution of the SME: it is the
Gaussian state closest to the actual conditioned state which may be
obtained by integrating the SME. This approach to the approximate solution of the SME appears in~\cite{Hal}. This is a more realistic estimator
for use in control than the SME since it only requires the integration of five
stochastic differential equations. The approximate solution is a
Gaussian mixed state which may be characterized by its mean position
$\langle x\rangle$ and momentum $\langle p\rangle$ and symmetric
second order moments $V_x,V_p,C$ the position and momentum variance
and the symmetric covariance $C = (1/2)\langle xp + px\rangle -
\langle x \rangle \langle p\rangle $ respectively.
\begin{eqnarray}
  \label{eq:csme}
  d\langle x\rangle  & = &  \langle p\rangle dt + 2\sqrt{2k} V_x dV
  \\ 
  d\langle p\rangle  & = &  -4B\langle x\rangle^3dt + 2A\langle
  x\rangle dt  - 12B\langle x\rangle V_x dt \nn \\ 
                     &   & + 2\sqrt{2k}C dV  + u dt, \\
  \dot{V}_x & = & 2 C - 8k V_x^2 \\
  \dot{V}_p & = & -24 B \langle x\rangle^2 C + 4 A C - 24 B C V_x \nn
  \\ 
            &   & + 2 (k+\beta) \hbar^2 - 8 k C^2 \\
  \dot{C} & = & V_p - 12 B \langle x\rangle^2 V_x + 2 A V_x \nn \\
          &   & - 12 B V_x^2 - 8 k C V_x
\end{eqnarray}
where $dV=dQ -\langle x\rangle dt$. Thus from an initial state the
observer may propagate this Gaussian estimate of the true conditioned
state given a particular measurement record. Note that since the full
SME is not in fact integrated the noise processes $dW$and $dV$ are not
the same. In our pure state trajectory simulations we perform the
stochastic integration of Eq.~(\ref{twellsme}) for different
realizations of the Wiener increments $dW$ that in turn determine, for
each trajectory, values of $dQ$ that are used to integrate the five
estimator equations. In order to obtain equations for pure states it
is also necessary to introduce a second Wiener increment to account
for the thermal noise as described in Section~\ref{sec:contqmeas}.

The state estimate is then used to determine the values of $u$. Under
linearized LQG control $u=u_1+u_2+u_3$ where
\begin{eqnarray}
   u_0 & = & 2 A \langle x\rangle - 4 B \langle x\rangle^3 \\
   u_1 & = & - \tilde{u} (\langle x\rangle - x_0) \\
   u_2 & = & - (\sqrt{2 \tilde{u} + \Gamma}) (\langle p\rangle - p_0)
   \\ 
   \tilde{u} & = & \partial_{\langle x \rangle} u_0 +
   \sqrt{[\partial_{\langle x \rangle} u_0]^2 + \Gamma}. 
\end{eqnarray}
The current target points in phase space are $x_0$ and $p_0$. Here
$\Gamma$ is a `free' parameter which one chooses to set the overall
strength of the feedback.

As a particular example we choose $A=2$ and $B=A/18$, which puts the
two minima at $\pm 3$, with a well depth of $13.5$. Since we set
$\hbar=1$, this puts the problem deep in the quantum regime, since the
potential varies considerably over the phase space area $\hbar$.
Because of this, the density (Wigner function) for the particle is
forced to be broad on the scale of the occupiable phase space, which
is a key limiting factor in the problem. We choose $\beta=0.1$, which
gives a thermal heating rate $d\langle E\rangle/dt=0.1$. Due to the
thermal heating, feedback control is essential to maintain a desired
behavior. In implementing the sub-optimal estimation and control
strategy described above, we have the choice of measurement strength
$k$ and feedback strength $\Gamma$. We find that it is possible to
obtain a fairly effective control with a choice of $k=0.3$ and
$\Gamma=100$.  A resulting trajectory for the system, given a target
position that switches between the well minima is shown in
figure~\ref{fig1}, along with the strength of the linear force applied
as a result of the control strategy. To evaluate the efficacy of the
control, we also plot the RMS deviation of the average position from
the target position, and plot this in Figure 2. We see from this that
the system achieves the target position within an average error of
$\pm 0.6$. When the target is switched, the system relaxes to the
desired value with a time constant of $\sim 3$.

\begin{figure}
\centerline{\psfig{file=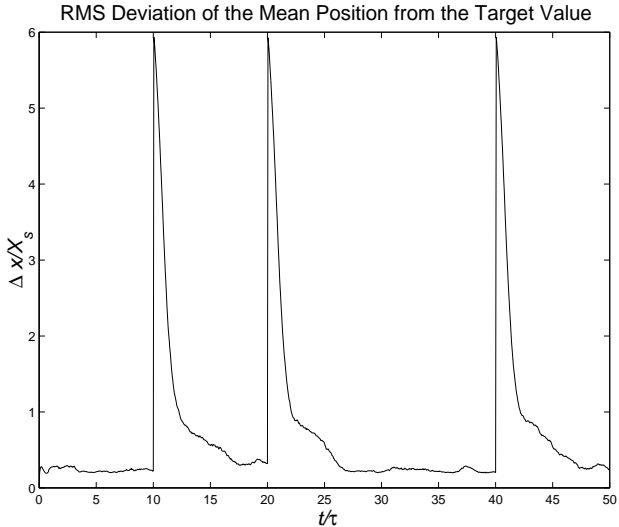,width=3.25in,height=2.8in}}
\caption{\narrowtext RMS deviation of the position from the target
value as a function of time. This was obtained by averaging over 1000
trajectories. The units are $X_{\mbox{\scriptsize s}} = 
\sqrt{\hbar/(m\nu)}$ and $\tau = 1/\nu$, where $m$ is the mass of the 
particle and $\nu$ is an arbitrary frequency. In the text we have 
set $\hbar = m = \nu = 1$, so that all quantities are dimensionless.}
\label{fig2}
\end{figure}

While this strategy is fairly effective, it is limited by specifically
quantum effects. In order to maintain a Gaussian state in the presence
of the non-linear potential the combined effect of the thermal noise
and measurement must be sufficiently strong, and this results in
unwanted heating which must be countered by the feedback. While this
is a limitation of the Gaussian estimator, there is still a more
fundamental limitation. In the presence of noise, the measurement must
be sufficiently strong in order to obtain sufficient information about
the system to control it. In this case we found we needed a
measurement strength three times that of the noise, resulting in the
corresponding heating. Naturally, these quantum limiting features are
ultimately due to the size of $\hbar$; as $\hbar$ decreases, the
measurement induced heating rate, as well as the rate at which the
Wigner function deforms from Gaussian, is reduced. It is to be
expected that with the use of more sophisticated estimation
techniques, and more subtle quantum control strategies, the simple
method we have outlined here can be beaten, possibly significantly,
and the development of such techniques constitutes a central problem
for future work in quantum feedback control.


\section{Conclusion}
In this paper we have argued that it is useful to consider quantum
feedback control in the light of methods developed in classical
control theory. In order to do this it is important to understand the
relationship between the two theories. We began by comparing the
formulations of these theories, in order to identify conceptual
analogies. We then considered three ways in which the quantum control
problem could be formally mapped to the classical problem, and
discussed if and when these formulations may be addressed directly
with the classical theory.

As an example, we applied the ideas presented here to the control of
the position of a single quantum particle in a non-linear potential
deep in the quantum regime. In this case we fixed both the measurement
observable (system/environment coupling) and the unraveling, and
considered the use of sub-optimal estimation and control
strategies. While this approach was fairly effective, it is clearly
limited by quantum effects.

As experimental techniques improve, and quantum technology becomes
increasingly relevant in practical applications, we can anticipate
that questions of quantum feedback control will become increasingly
important. It is clear that most questions regarding the optimal
observables, unravelings, and control strategies required for quantum
feedback control problems, and the effectiveness of sub-optimal
estimation algorithms, are as yet unanswered, and that this field
presents a considerable theoretical challenge for future work.

\section{acknowledgements}
SH, KJ and HM would like to thank Tanmoy Bhattacharya, Chris Fuchs and
Howard Barnum for helpful discussions. This research was performed in
part using the resources located at the Advanced Computing Laboratory
of Los Alamos National Laboratory.

\end{multicols}
\widetext
\begin{multicols}{2}

\end{multicols}

\end{document}